\documentclass[%
 aip,
 jmp,%
 amsmath,amssymb,
reprint,%
]{revtex4-1}

\usepackage{graphicx}
\usepackage{dcolumn}
\usepackage{bm}

\begin{document}


\title[Consideration of Covariant Quantization of Electromagnetic Field]{Consideration of Covariant Quantization of Electromagnetic Field}

\author{Masahito Morimoto}


\date{\today}

\begin{abstract}
We examine a covariant quantization of electromagnetic fields by using an operator derived from a constant scalar that can be called extended Lorentz gauge. The quantization can avoid an inconsistency between Lorentz gauge and a commutation relation, which can eliminate the need for introduction of physical state defined by a subsidiary condition and auxiliary field in Lagrangian density in Lorentz gauge. By using this quantization and indefinite metric straightforwardly, all quantum phenomena can be provided without enigmatic and paradoxical ``probability interpretation''. 

%
\end{abstract}

\maketitle
\section{Introduction}
In quantum theory, some paradoxes have been acknowledged, which are associated with relativity typified by ``Schr\"odinger's cat'' and ``Einstein, Podolsky and Rosen (EPR)'' \cite{Schr-cat,EPR}. In order to solve the paradoxes, the author has reported the alternative interpretation for quantum theory utilizing quantum field formalism with unobservable potentials \cite{Morimoto1} that can be identified as ``hidden variables'' similar to Aharonov-Bohm effect \cite{ABeffe,Tonomura1,Tonomura2} and rigorous mathematical treatment using tensor form in keeping with the local representation, i.e., consistent with relativity. The interpretation can omit the quantum paradoxes and be applicable to elimination of infinite zero-point energy, spontaneous symmetry breaking, mass acquire mechanism, non-Abelian gauge fields and neutrino oscillation, which can lead to the comprehensive theory.

In addition, the author also shows the existence of the unobservable potentials can explain the quantum eraser and delayed choice experiment \cite{Doitysqe,clauser1974experimental,PhysRevLett.84.1,PhysRevA.65.033818}, and the interference between photons and the unobservable potentials violates Bell's inequalities \cite{Bell,aspect1981experimental,aspect1982experimental-2,PhysRevLett.100.220404} in keeping with the local representation, which is consistent with relativity \cite{Morimoto2}.

However, a sensitive mathematical formalism for quantization procedure has not been described in the author's report \cite{Morimoto1, Morimoto2}. 

In this letter, the covariant quantization procedure of electromagnetic field in Lorentz gauge without subsidiary conditions \cite{0370-1298-63-7-301,Bleuler} and auxiliary field is discussed. 

\section{Traditional Issue in the Lorentz gauge}
First, we present a general outline of a covariant quantization procedure of electromagnetic field in Lorentz gauge and need for introduction of auxiliary field and physical state defined by a subsidiary condition briefly.

Let consider following Maxwell equations in free space.
\begin{eqnarray}
\left( \Delta -\frac{1}{c^{2}} \frac{\partial ^{2}}{\partial t^{2}} \right) {\bf A}- \nabla \left( \nabla \cdot {\bf A} + \frac{1}{c^{2}} \frac{\partial \phi}{\partial t} \right) = 0 \nonumber \\
\left( \Delta -\frac{1}{c^{2}} \frac{\partial ^{2}}{\partial t^{2}} \right) \phi + \frac{\partial}{\partial t} \left( \nabla \cdot {\bf A} + \frac{1}{c^{2}} \frac{\partial \phi}{\partial t} \right) = 0
\label{eq:Maxi}
\end{eqnarray}
where $ {\bf A} $ and $ \phi $ are vector and scalar potentials respectively. By using four-vector
\begin{equation}
A^{\mu} = (A^{0}, \ A^{1}, \ A^{2}, \ A^{3} ) = (\phi /c, \ {\bf A})
\end{equation}
and $ \partial_{\mu} = ( 1/c\partial t , \ 1/\partial x , \ 1/\partial y , \ 1/\partial z ) \equiv ( 1/\partial x^{0} , \ 1/\partial x^{1} , \ 1/\partial x^{2} , \ 1/\partial x^{3} ) $, we can describe the Maxwell equation as following covariant form.
\begin{eqnarray}
\Box A^{\nu} - \partial_{\mu} \partial^{\nu} A^{\mu} & = & 0
\label{eq:Maxwell}
\end{eqnarray}
Where
\begin{eqnarray}
A_{\mu} = \texttt{g}_{\mu \nu} A^{\nu} & , & A^{\mu} = \texttt{g}^{\mu \nu} A_{\nu} \nonumber \\
\partial_{\mu} = \texttt{g}_{\mu \nu} \partial^{\nu} & , & \partial^{\mu} = \texttt{g}^{\mu \nu} \partial_{\nu} \nonumber \\
 \texttt{g}_{\mu \nu} = \texttt{g}^{\mu \nu} & = & \left[ 
 \begin{array}{cccc}
 1 & 0 & 0 & 0 \\
0 & -1 & 0 & 0 \\
0 & 0 & -1 & 0 \\
0 & 0 & 0 & -1\\
 \end{array} 
\right]
\end{eqnarray}
In order to adopt canonical quantization, the following classical lagrangian density has been introduced.
\begin{eqnarray}
\pounds_{class} & = & -\frac{1}{4}F_{\mu \nu}F^{\mu \nu} \equiv -\frac{1}{4} (\partial_{\mu} A_{\nu} - \partial_{\nu} A_{\mu}) (\partial^{\mu} A^{\nu} - \partial^{\nu} A^{\mu}) \nonumber \\
& = & -\frac{1}{2} \partial_{\mu} A_{\nu} \partial^{\mu} A^{\nu} + \frac{1}{2} \partial_{\mu} A_{\nu} \partial^{\nu} A^{\mu}
\label{eq:Lag0}
\end{eqnarray}
Indeed the following Euler-Lagrange equation gives Maxwell's equations (\ref{eq:Maxwell}).
\begin{equation}
\partial_{\mu} \frac{\partial \pounds_{class}}{\partial (\partial_{\mu} A_{\nu})} - \frac{\partial \pounds_{class}}{\partial A_{\nu}} =0
\label{eq:EL}
\end{equation}
By using this lagrangian density, the canonically conjugate variables $ \pi^{i}, \ i = {\rm (1, \ 2, \ 3)} $ can be defined as follows.
\begin{eqnarray}
\pi^{i} & = & \frac{\partial \pounds_{class}}{\partial \dot{A}_{i}} = -\frac{1}{4} \frac{\partial}{\partial \dot{A}_{i}} (F_{0 i} F^{0 i}) \nonumber \\
& = & -\frac{1}{4} \frac{\partial}{\partial \dot{A}_{i}} ((\partial_{0} A_{i} - \partial_{i} A_{0}) (\partial^{0} A^{i} - \partial^{i} A^{0}) \nonumber \\
& & + (\partial_{i} A_{0} - \partial_{0} A_{i}) (\partial^{i} A^{0} - \partial^{0} A^{i}))\nonumber \\
& = & \partial^{i} A^{0} -\partial^{0} A^{i}
\end{eqnarray}
However the conjugate variable $ \pi^{0} $ can not be defined as follows.

\begin{equation}
\pi^{0} = \frac{\partial \pounds_{class}}{\partial \dot{A}_{0}} = 0
\label{eq:pi=0}
\end{equation}
Therefore a number of fixing gauge conditions have been proposed. Well-known gauges are Coulomb gauge $ \nabla \cdot {\bf A} = 0 $ and Lorentz gauge $ \partial_{\mu} A^{\mu} = 0 $. Because Coulomb gauge spoils the explicit covariance due to the
separation of $ A_{0} $ from the four-vector, fixing the Lorentz gauge has been examined by using following lagrangian density.
\begin{equation}
\pounds_{0} = -\frac{1}{4}F_{\mu \nu}F^{\mu \nu} -\frac{1}{2} (\partial_{\rho} A^{\rho} )^2
\label{eq:lag0}
\end{equation}  
When we use the lagrangian density (\ref{eq:lag0}), Maxwell's equations in Lorentz gauge can be obtained from Euler-Lagrange equation (\ref{eq:EL}).
\begin{eqnarray}
\Box A^{\nu} & = & 0
\label{eq:Simp-Maxwell}
\end{eqnarray}
Here the action integral of  (\ref{eq:Lag0}) is as follows.
\begin{eqnarray}
S & \equiv & \int d^4 x \pounds_{class} \nonumber \\
& = & \int d^4 x (-\frac{1}{2} \partial_{\mu} A_{\nu} \partial^{\mu} A^{\nu} + \frac{1}{2} \partial_{\mu} A_{\nu} \partial^{\nu} A^{\mu})
\end{eqnarray}
The second term of the above integral is calculated to be $ \frac{1}{2} ( \partial_{\mu} A^{\mu} )^2 $ by partial integration. Then the lagrangian density (\ref{eq:lag0}) which derives Maxwell's equations (\ref{eq:Simp-Maxwell}) can be calculated to be following lagrangian density.
\begin{equation}
\pounds'_{0} = -\frac{1}{2} \partial_{\mu} A_{\nu} \partial^{\mu} A^{\nu}
\label{eq:lag2}
\end{equation}  
The canonically conjugate variables can be obtained by using this lagrangian density as follows .
\begin{eqnarray}
\pi^{\mu} = \frac{\partial \pounds'_{0}}{\partial \dot{A}_{\mu}} & = & - \dot{A}^{\mu}
\label{eq:CCV}
\end{eqnarray}
Here quantization is performed by replacing the fields with operators and set the following equal-time commutation relations.
\begin{eqnarray}
[ A^{\mu} ({\bf x}, t), \pi^{\nu} ({\bf x}', t) ] & = & - [ A^{\mu} ({\bf x}, t), \dot{A}^{\nu} ({\bf x}', t) ] \nonumber \\
& = & i \texttt{g}^{\mu \nu} \delta^3 ({\bf x}-{\bf x}')
\label{eq:comrel}
\end{eqnarray}
\begin{eqnarray}
[ A^{\mu} ({\bf x}, t), A^{\nu} ({\bf x}', t) ] = [ \pi^{\mu} ({\bf x}, t), \pi^{\nu} ({\bf x}', t) ]  = 0
\label{eq:comrel2}
\end{eqnarray}
However (\ref{eq:comrel}) and (\ref{eq:comrel2}) derive the following relations.
\begin{equation}
[ \partial_{\mu} A^{\mu} ({\bf x}, t), A^{\nu} ({\bf x}', t) ] = i \texttt{g}^{0 \nu} \delta^3 ({\bf x}-{\bf x}') \neq 0
\label{eq:bibunn-comrel}
\end{equation}
Hence (\ref{eq:bibunn-comrel}) is inconsistent with Lorentz gauge $ \partial_{\mu} A^{\mu} = 0 $ as an operator.

Therefore some other lagrangian densities have been proposed. The following lagrangian with auxiliary scalar field $ B $, which is called Nakanishi-Lautrup formalism, will be the most comprehensive form \cite{doi:10.1143/PTPS.51.1}.
\begin{equation}
\pounds_{NL} = -\frac{1}{4}F_{\mu \nu}F^{\mu \nu} + B \partial^{\mu} A_{\mu} + \frac{1}{2} \alpha B^2
\label{eq:NakaLaut}
\end{equation}
Where $ \alpha $ is an arbitrarily real parameter. The inconsistency between Lorentz gauge $ \partial_{\mu} A^{\mu} = 0 $  and (\ref{eq:bibunn-comrel}) can be avoided by using the lagrangian (\ref{eq:NakaLaut}), introduction of physical states $ | {\rm phys} \rangle $ and a restriction of Lorentz gauge in terms of the physical states defined by a subsidiary condition, i.e., $ \langle {\rm phys} | \partial _{\mu} A^{\mu} | {\rm phys} \rangle = 0 $.

\section{Extended Lorentz gauge}
The approach using (\ref{eq:NakaLaut}) seems to be an artificially imposed mathematical technique by introducing an unreal physical field $ B $ and unphysical man-made mathematical formality called ``subsidiary condition''. In addition, the approach has been introduced for avoidance of negative norm as premises for ``probability interpretation''. However the author has been discussed in reference\cite{Morimoto1, Morimoto2} that the negative norm is indispensable in real nature (reality) and the ``probability interpretation'' is not justified in real nature except only for mixed states, i.e., statistical sense. 

 In this section, we discuss the lagrangian density (\ref{eq:lag2}) and Maxwell equation in Lorentz gauge (\ref{eq:Simp-Maxwell}) again. It is to be noted that Lorentz gauge is dispensable for deriving (\ref{eq:Simp-Maxwell}). Indeed (\ref{eq:Simp-Maxwell}) is derived from lagrangian density (\ref{eq:lag0}) or (\ref{eq:lag2}) independently of Lorentz gauge.
Alternatively the following condition is indispensable from (\ref{eq:Maxwell}).
\begin{equation}
\partial_{\mu} \partial^{\nu} A^{\mu} = 0
\label{eq:newgage}
\end{equation}
Hence from Lorentz invariance 
\begin{equation}
\partial_{\mu} A^{\mu} = \epsilon ({\rm scaler})
\label{eq:LI1}
\end{equation}
This condition (we call this ``extended Lorentz gauge'') also has the following gauge invariance of (\ref{eq:Maxwell}) by introducing an arbitrary scalar function $ \chi $.
\begin{equation}
A'^{\mu}  = A^{\mu} + \partial^{\mu} \chi
\end{equation}
Although this replacement has been well known, we can imagine that four-vectors $ A^{\mu} $ move on the bias vectors $ \partial^{\mu} \chi $ such as signal components of an operational amplifier in an electric circuit or surface wave of the sea. 

By choosing $ \Box \chi = 0 $, (\ref{eq:LI1}) can be obtained repeatedly as follows. 
\begin{equation}
\partial_{\mu} A'^{\mu} = \partial_{\mu} A^{\mu} + \Box \chi = \partial_{\mu} A^{\mu} = \epsilon
\end{equation} 
Hence
\begin{equation}
A^{\mu} = A_{L}^{\mu} + f^{\mu}
\label{eq:Asol}
\end{equation}
where $A_{L}^{\mu} $and $ f^{\mu} = f^{\mu}({\bf x}, t) $ are a general solution of Lorentz gauge, i.e., $ \partial_{\mu} A_{L}^{\mu} = 0 $, and a linear formula as a function of $ {\bf x}, t $ with $ \partial_{\mu} f^{\mu} = \epsilon $ respectively. The most common linear formula $ f^{\mu} $ is the same as a coordinate transformation in form described as follows.
\begin{equation}
f^{\mu} = \beta ( a^{\mu}_{ \nu} x^{\nu} + b^{\mu} )
\end{equation}
where $ \beta $ is a constant for fixing the appropriate dimension. Therefore 
\begin{equation}
\epsilon = \beta ( a^{0}_{0} + a^{1}_{1} + a^{2}_{2} + a^{3}_{3} ) = \mathrm{Tr} (\varepsilon)
\label{eq:Trace}
\end{equation}
Where, $\varepsilon $ is $4\times 4$ matrix with matrix elements $a^{\mu}_{\nu}$ multiplied by $\beta$.
Here we replace $\epsilon$ with operator $ \hat{\epsilon} $ and substitute for (\ref{eq:bibunn-comrel})
\begin{equation}
[ \partial_{\mu} A^{\mu} ({\bf x}, t), A^{\nu} ({\bf x}', t) ] = [ \hat{\epsilon} , A^{\nu} ({\bf x}', t) ] = i \texttt{g}^{0 \nu} \delta^3 ({\bf x}-{\bf x}') \neq 0
\label{eq:bibunn-comrel-2}
\end{equation}

\section*{Discussion}
We examine the commutation relation $  [ \partial_{\mu} A^{\mu} , A] = [ \hat{\epsilon} , A] \equiv \hat{\epsilon} A - A \hat{\epsilon} \neq 0 $ by using matrix representation of the operator.
The matrix representation of the photon creation and annihilation operators, $ A^{\dagger} $ and $ A $, and photon number state vectors $ | {\bf n} \rangle $ are expressed as follows.

\begin{equation}
{\bf A} = \left[ 
 \begin{array}{rrrrr}
0 &1 & 0 & 0 & \cdots \\
0 & 0 & \sqrt{2} & 0 & \cdots \\
0 & 0 & 0 & \sqrt{3} & \cdots \\
0 & 0 & 0 & 0 & \cdots \\
\vdots & \vdots & \vdots & \vdots & \ddots \\
\end{array} 
\right], \ \
 {\bf A^{\dagger}} = \left[
 \begin{array}{rrrrr}
0 &0 & 0 & 0 &\cdots \\
1 & 0 & 0 & 0 & \cdots \\
0 & \sqrt{2} & 0 & 0 & \cdots \\
0 & 0 & \sqrt{3} & 0 &  \cdots \\
\vdots & \vdots & \vdots & \vdots & \ddots \\
\end{array} 
\right] 
\end{equation}

\begin{equation}
| 0 \rangle = \left[
 \begin{array}{c}
1 \\
0 \\
0 \\
0 \\
\vdots \\
\end{array} 
\right], 
| 1 \rangle = \left[
 \begin{array}{c}
0 \\
1 \\
0 \\
0 \\
\vdots \\
\end{array} 
\right],
| 2 \rangle = \left[
 \begin{array}{c}
0 \\
0 \\
1 \\
0 \\
\vdots \\
\end{array} 
\right],
| 3 \rangle = \left[
 \begin{array}{c}
0 \\
0 \\
0 \\
1 \\
\vdots \\
\end{array} 
\right], \cdots
\end{equation}

When $ \hat{\epsilon} = \epsilon I $, where $ I $ is an identity matrix or operator, the operator $ \hat{\epsilon} $ just serves as the constant scalar $ [ \hat{\epsilon} , A] = \epsilon [ I , A] = \epsilon ( I A - A I ) = 0 $,  then （\ref{eq:bibunn-comrel-2}） can not be obtained.
From (\ref{eq:Trace}), we should adopt the operator $\hat{\varepsilon}$ that satisfies $ \mathrm{Tr} ( \hat{\varepsilon} ) = \Sigma \varepsilon^{ii} = \epsilon $ when we replace $\epsilon$ with the operator $ \hat{\epsilon} $. Hence, the following matrix representation can be adopted as $\hat{\epsilon}$, where we set all diagonal elements to 0 for simplicity. Even if the diagonal element is not 0, the same result as below can be obtained.
\begin{equation}
\hat{\epsilon} = \left[
 \begin{array}{ccccc}
\varepsilon^{00} & 0 & 0 & 0 & \cdots \\
0 & \varepsilon^{11} & 0 & 0 & \cdots \\
0 & 0 & \varepsilon^{22} & 0 & \cdots \\
0 & 0 & 0 & \varepsilon^{33} & \cdots \\
\vdots & \vdots & \vdots & \vdots & \ddots \\
\label{eq:epsilon}
\end{array} 
\right] 
\end{equation}
Here we can calculate as follows.
\begin{eqnarray}
[ \hat{\epsilon} , A] & \equiv & \hat{\epsilon} A - A \hat{\epsilon} \nonumber \\
& = & \left[
 \begin{array}{ccccc}
0 & \varepsilon^{00}  & 0 & 0 & \cdots \\
0 & 0  & \sqrt{2} \varepsilon^{11} & 0 & \cdots \\
0 & 0 & 0 & \sqrt{3} \varepsilon^{22} & \cdots \\
0 & 0  & 0 & 0 & \cdots \\
\vdots & \vdots & \vdots & \vdots & \ddots \\
\end{array} 
\right]  \nonumber \\
& - & \left[ 
 \begin{array}{ccccc}
0 & \varepsilon^{11}  & 0 & 0 & \cdots \\
0 & 0  & \sqrt{2} \varepsilon^{22} & 0 & \cdots \\
0 & 0 & 0 & \sqrt{3} \varepsilon^{33} & \cdots \\
0 & 0  & 0 & 0 & \cdots \\
\vdots & \vdots & \vdots & \vdots & \ddots \\
\end{array} 
\right]  \nonumber
\end{eqnarray}
\begin{eqnarray}
 & = & \left[
 \begin{array}{ccccc}
0 & \varepsilon^{00} -  \varepsilon^{11} & 0 & 0 & \cdots \\
0 & 0  & \sqrt{2} ( \varepsilon^{11} - \varepsilon^{22} ) & 0 & \cdots \\
0 & 0 & 0 & \sqrt{3} ( \varepsilon^{33} - \varepsilon^{22} ) & \cdots \\
0 & 0  & 0 & 0 & \cdots \\
\vdots & \vdots & \vdots & \ddots \\
\end{array} 
\right] \nonumber \\
\end{eqnarray}
Hence if at least one $ \varepsilon^{ii} (i >0) $ is $ \varepsilon^{ii} \neq \varepsilon^{i\pm1 i\pm1} $ then $ [\hat{\epsilon}, A] \neq 0 $ will be satisfied.

By utilizing the relationship (\ref{eq:Trace}), we can define the operator $ \hat{0} \equiv \hat{\epsilon} $ that satisfies $ \mathrm{Tr} ( \hat{\varepsilon} ) = \Sigma \varepsilon^{ii} = 0 $ because $ \hat{\epsilon} $ satisfies $ \mathrm{Tr} ( \hat{\varepsilon} ) = \Sigma \varepsilon^{ii} = \epsilon $ like (\ref{eq:epsilon}). Therefore $ \hat{0} \equiv \partial_{\mu} A^{\mu} ({\bf x}, t) $ satisfies (\ref{eq:bibunn-comrel}) as the conventional Lorentz gauge  

Since classical Lagrangian density (\ref{eq:Lag0}) assumes free space which is an ideal vacuum with no geometry\cite{Morimoto1}, the same approach will be applicable by introducing a phase into $A_{0}$ that includes the existence of the geometry.

\section{Conclusions}
We have presented a considerable potential for resolving the difficulty raised in a covariant quantization of electromagnetic fields in Lorentz gauge by examining the operators derived from constant scalar. The presented quantization procedure does not need subsidiary condition, physical states and auxiliary field, which seems to be physically unnatural. If the description discussed in this paper by using the constant scalar operator is valid, we can consider Lorentz gauge becomes $ \partial_{\mu} A^{\mu} \neq 0 $ as an operator. Therefore Lorentz gauge can be used without inconsistency between Lorentz gauge and commutation relation.



\end{document}